\input psfig.tex

\documentstyle{article}
\begin{document}

\thispagestyle{empty}
\begin{tabbing}
\hskip 7.0in \= {\makebox[0in][r] {Imperial/TP/94-95/36} }
\\
\hskip 7.0in \= {\makebox[0in][r] {Submitted to Mon. Not. R. Astron. Soc.}}
\\
\end{tabbing}
\vskip 1cm
\begin{center}
{\Large\bf A Statistic for identifying cosmic string wakes and other
sheet-like structure}
\vskip 1.2cm
{\large\bf James Robinson and Andreas Albrecht}\\
Blackett Laboratory, Imperial College\\
Prince Consort Road, London SW7 2BZ  U.K.\\
\end{center}
\vskip 1cm

\begin{abstract}
We describe an implementation of the structure functions of Babul
\& Starkman \cite{s123}, in order to quantify the 
``sheet-like'' nature of a distribution of matter.
We test this statistic on a toy model describing cosmic string
wakes, and show that it does a better job than other statistics
which have been proposed for distinguishing non-Gaussianity in the
form of sheets.
We conclude that  the most favoured cosmic string model is unlikely to
produce a significant increase in the sheet-like nature of the matter
distribution beyond that which occurs in Gaussian models (with the same
power spectrum) due to the
formation of Zeldovich pancakes.
Although the statistic was developed in the context of  cosmic string
wake formation,  we  expect it to be useful for
comparing the observed galaxy distribution with a wide range of
theoretical  models with different power spectra.
\end{abstract}

\section{Introduction}

Advances in astronomy and theoretical physics are providing an
increasingly detailed picture of  the matter distribution in the
universe, and of the predictions made  by various
theoretical models.  Comparisons of theory with observations are made
using statistics (such as the two-point correlation function) which
might be  able to identify significant differences or similarities
among the models and data.
Clearly, making the most of these comparisons involves identifying
which are the
most significant statistics to calculate.

Many models of
primordial perturbations are ``Gaussian'', and the density field at
early times is completely characterized by the two-point correlation
function (or power spectrum).  However, on many scales we observe the
matter distribution after the simple connections between the two point
correlation function and other measurements have been destroyed by
non-linear evolution.

The observed galaxy distribution exhibits many sheet-like features. So
do essentially all of the theoretical models. Sheet-like behaviour
is expected around scales  where gravity is
just going non-linear.  A typical asymmetrical perturbation will
collapse more quickly in one direction than the others, lending a
sheet-like nature to the distribution of matter.  In Gaussian models,
the resulting sheet-like properties  are a  reflection of  features
of the initial power spectrum. It may well be that statistics sensitive to
sheets are the best tool for exposing these features.

Wakes from cosmic strings can also produce sheet-like perturbations in
the matter distribution.  It is not yet clear whether in realistic cosmic
string scenarios such wakes have a significant observational impact,
or whether they are washed out by various competing factors.
A statistic which is as sensitive as possible to sheet-like features
will enhance our ability to address this question.

For this work, we constructed a toy model based on the cosmic string
picture.  We produced realizations of the matter distribution
today based on a purely Gaussian set of primordial perturbations, and
compared the results with the case where an individual cosmic string
wake was added in ``by hand''.  By  studying this toy model, we
found an implementation of the structure functions of Babul \&
Starkman \cite{s123} which was better at identifying non-Gaussianity
in the form of sheets
than any other statistic we tried\footnote{In an earlier version of
this paper we found that a different measure of flatness
was best at identifying wakes in our toy model. Since then
we have developed a parallel version of our computer
code, which has allowed us to scan a wider range of the parameter space
available to the original statistic. Contrary to our original intuition, we
have found that the wakes signal can be made even stronger by
measuring the flatness of sections of the matter distribution which
are several times smaller than the curvature scale of the wake. We find
that these improvements have brought our statistic more in line with
the use of the structure functions described by Babul \& Starkman in
\cite{s123}.}. We stress that although the cosmic
string picture played a role in the development of this statistic,
we expect its utility to extend beyond the cosmic string picture.

The paper is organized as follows:  The following section sketches
our toy model and gives a qualitative discussion of the nature of the
problem.   Then Section~\ref{sec-statistics} describes our
implementation of the structure functions of Babul \& Starkman,
illustrates
its application to the toy model, and compares it with other
statistics.
Section~\ref{sec-conclusion} gives
our conclusions.
Appendix~\ref{sec-model} describes the toy model in detail.
Appendix~\ref{sec-visibility}
describes a new
``velocity coherence'' criterion for the visibility of cosmic string
wakes, and Appendix~\ref{sec-shell}
describes how various two dimensional structures will show up in this
statistic.

\section{Qualitative look at our model and its output}

\label{sec-qualitative}
\subsection{The Toy Model}
For the work described in this paper, we wanted a set of physically
motivated models for the origin of large scale structure, which
exhibit sheet-like properties in various degrees.
Just such a set of models is described by Albrecht \& Stebbins
\cite{HDM}. They consider density perturbations
induced by three models of cosmic
string (AT, I and X) in a background of HDM.
To see how ``wakey'' the resulting perturbations are, they propose
the following criterion: They ask whether the
dominant contribution to the linear density contrast in a sphere of radius $R$
centered on a wake comes from the wake ($\Delta_{\rm W}$) or from the
Gaussian noise of other perturbations (characterized by an RMS value
$\Delta_{\rm rms}$). The wake may be said to stand out if $\Delta_{\rm W}
\succeq \Delta_{\rm
rms}$. In particular, they consider ``maximal''
wakes, which give the largest single contribution to linear
density perturbations on a scale of $1$Mpc.
According to their criterion, maximal wakes in the X model stand out
strongly, those in the I model are on the borderline, and those in the
AT model do not stand out.

Our toy model works as follows: We
consider the
initial density
perturbations induced in a box of side $L$ to be made up of two
components. Firstly, we take a realization of a Gaussian random
field with a cosmic string power spectrum, as worked out by Albrecht
\& Stebbins \cite{HDM}.
Secondly, we
add in the perturbation induced by a bit of string which enters the box
when the coherence length of the string network is equal to the
box size.
Such a bit of string is approximately straight, and moves in an
approximately straight line. We can therefore imagine it to seed a
perturbation in a flat plane passing through the middle of the box.
Having worked out the initial
density field in our box, we evolve it to the present day using
the adhesion approximation \cite{adhesion}. For consistency with the
results presented for the three wakes models in \cite{HDM}, we
use $h=1.0$ throughout this paper.
In Appendix~\ref{sec-model} we describe our toy model in detail, and
argue that it can give a fairly realistic picture of what wakes look
like today.

\subsection{Eyeball results}
\label{sec-eyeball}
We have used our toy model to study maximal wakes (as discussed in the
previous subsection) in the AT, I and X models. We chose to
investigate these wakes as they also give the largest single
contribution to the linear density contrast on a range of scales up to
$\sim 20$Mpc, and consequently have
the best chance of being observed in the universe today. We note that the time
these maximal wakes are laid down ($\eta=6 \eta_{\rm eq}$) is
independent of the network model we consider. Instead, it is a
generic property of the HDM we are perturbing\footnote{This fact has been
employed in the past to use maximal wakes to model the entire impact
of cosmic strings -- see for instance Perivolaropoulos, Brandenberger \&
Stebbins \cite{times}. However, this approach can only be valid if the maximal
wakes stand out.}.

Results for boxes perturbed by wakes in each model
are presented in a form suitable for
examination by eye in
Figure~\ref{fig-eyeballall}.
Each picture shows a side view of all
particles in the box, with the orientation chosen so that the plane
which has been perturbed by the wake is seen end on, and occupies a
vertical line through the middle of the picture.
(These pictures were generated using $32^3$ particles evolved on a $128^3$
lattice). For comparison, we also show boxes
evolved from the same
Gaussian initial conditions, but where a
wake has not been added. The boxes are 10Mpc, 20Mpc and 60Mpc
respectively for the
AT, I and X models\footnote{Due to peculiarities in the
``extreme'' X model, the natural boxsize to choose is 60MPC,
approximately $\frac{2}{3}$ of the coherence scale.}. The pictures for
each model look similar
because they are generated using the same random numbers. We see that
only the X wake stands out
clearly. In the I and the AT cases, there are new features in the
boxes containing wakes. However, these features do not seem to be
intrinsically different from the type of things which appear when the
Gaussian alone undergoes collapse, and they certainly do not dominate
the resulting density field.

\begin{figure*}
\begin{minipage}{8.2in}
\setlength{\unitlength}{1in}
\begin{picture}(8.2,9.0)
\end{picture}
\caption{AT, I and X boxes (top to bottom) with and
without a wake (left to right).}
\label{fig-eyeballall}
\end{minipage}
\end{figure*}

There are two distinct reasons why X wakes stand out more than
I and the AT wakes.
\begin{enumerate}
\item {\em Properties of the initial density fields:}
X wakes stand out more strongly in the initial density
fields. According to the Albrecht and Stebbins wakiness criterion, both
I and X wakes should stand out, but X wakes should stand out more. In
Appendix~\ref{sec-visibility} we propose a new
visibility criterion for string wakes based on {\em velocity
coherence} in initial density fields. Using this
criterion, we demonstrate that there is not just a quantitative, but a
{\em qualitative} difference
between the manner in which X and I wakes stand out: X wakes dominate
the bulk motions of matter, while I wakes do not. This prediction is
borne out by an inspection of the pictures in this section.

\item {\em Effects of the non-linear evolution:}
The degree of non-linear evolution in each box affects the
visibility of the
wakes.
As we go from AT through I to X, we are looking at larger and larger
boxes, because the coherence scale of the wakes at $\eta = 6 \eta_{\rm
eq}$ is increasing. Consequently, we are looking at evolution which is
less and less non-linear.
Highly non-linear evolution will cause the
wakes to break up into smaller objects, making them
less continuous and harder to observe. Further, it produces other
non-linear structures, such as Zeldovich pancakes, which provide
competition against which the wakes may have difficulty standing out.
So as we go
from AT through I to X, the
decrease in non-linearity of the evolution  tends to increase the
ease with which the wakes stand out.
\end{enumerate}

The combination of these two effects allows  the three cosmic string
models to give  an interesting range
of cases to analyze. Since the I model is the best motivated
by modern numerical simulations, the visual analysis  suggests
that individual string wakes will not
play a dominant role in realistic string scenarios.
We now turn to the problem of quantifying this issue.

\section{Statistics}
\label{sec-statistics}
\subsection{Counts in cells and the genus curve}
\label{sec-cic}
The pictures in Figure~\ref{fig-eyeballall} illustrate that maximal
wakes laid down in
the X model grow into significant structures today, and that those in
the I and AT model have a small influence on matter distributions
today. In this section we discuss how the resulting structures show up
in {\em counts in cells} \cite{salsaw} and the
{\em genus curve} \cite{gott1}, \cite{gott2}, both of which have
been put forward as useful measures for identifying non-Gaussianity
in the form of sheets \cite{brandenberger}.
In order to test these statistics,
we compare their
values for boxes evolved from the two types of initial
conditions discussed earlier:
\begin{enumerate}
\item
\label{box1}
Realization of a strings power spectrum, plus a single string wake (as
described in Appendix \ref{sec-model}).
\item
\label{box2}
Realization of a strings power spectrum on its own.
\end {enumerate}
For the purposes of this paper, we only consider the idealized case
where we have a well sampled map of the true density field. Issues
concerning finite sampling effects and redshift-space distortion
associated with real galaxy catalogues will be addressed in future work.

\begin{figure}
\centerline{\psfig{file=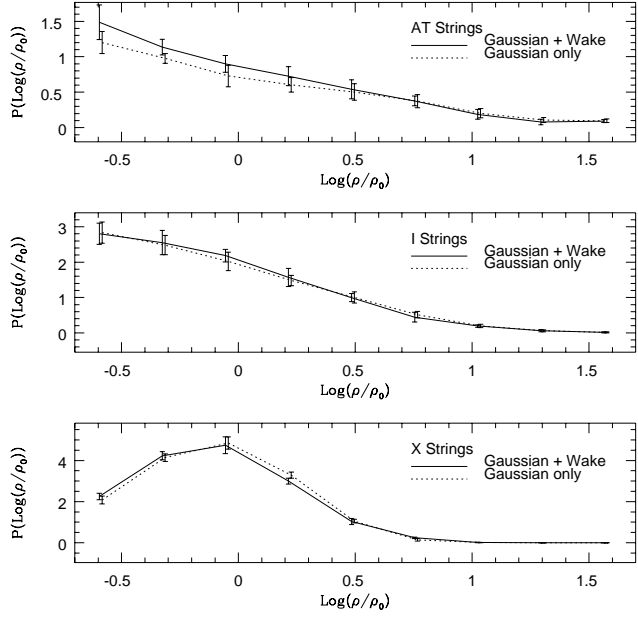,width=3.5in}}
\caption{Counts in cells distributions for X, I and AT boxes with and
without wakes. The boxes are $60$Mpc, $20$Mpc and $10$Mpc respectively. In
each case, the cell size is equal to the box size divided by $8$.}
\label{fig-cic8}
\end{figure}

\begin{figure}
\centerline{\psfig{file=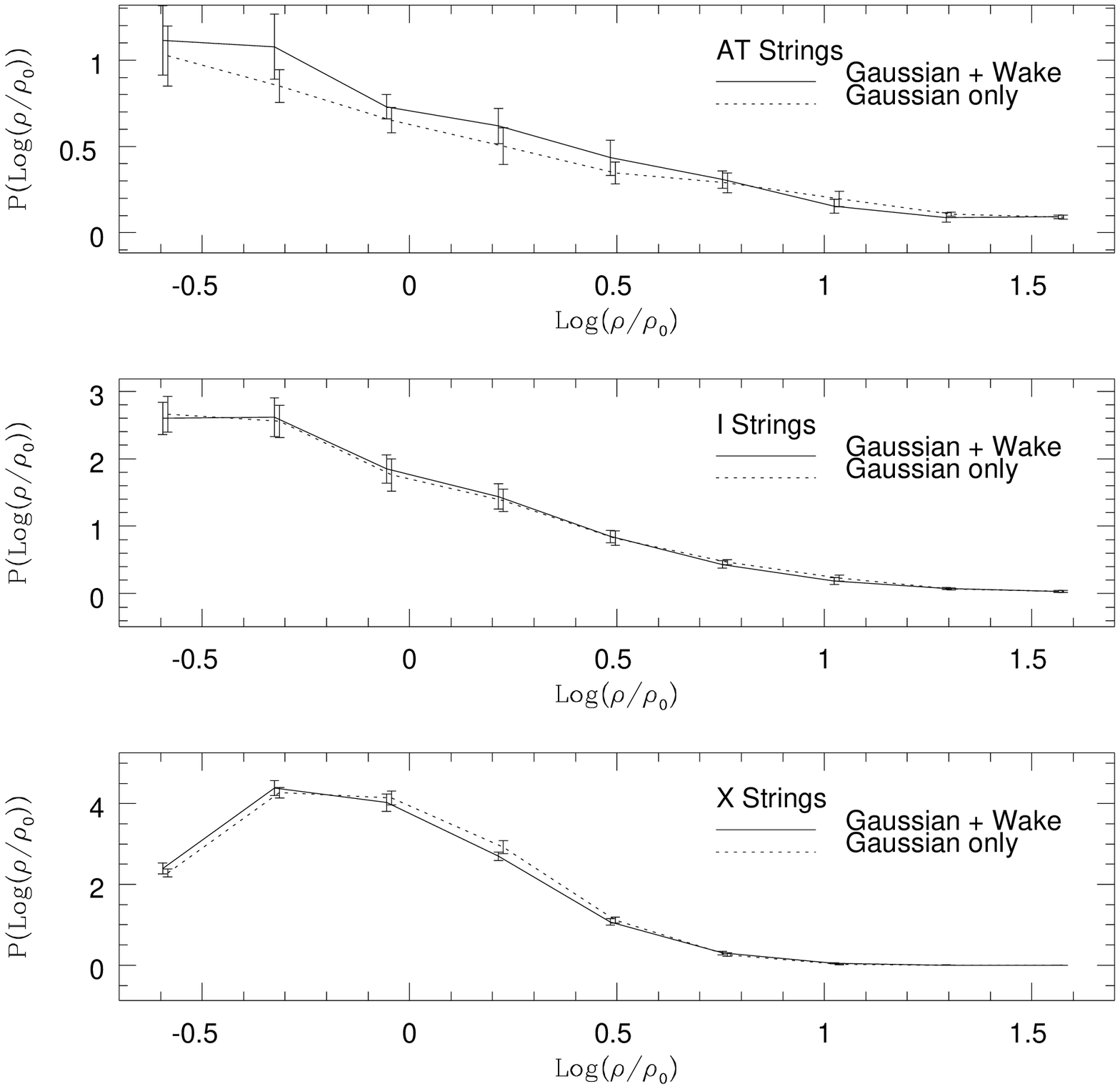,width=3.5in}}
\caption{Counts in cells distributions for X, I and AT boxes with and
without wakes. The boxes are $60$Mpc, $20$Mpc and $10$Mpc respectively. In
each case, the cell size is equal to the box size divided by $12$.}
\label{fig-cic12}
\end{figure}

\begin{figure}
\centerline{\psfig{file=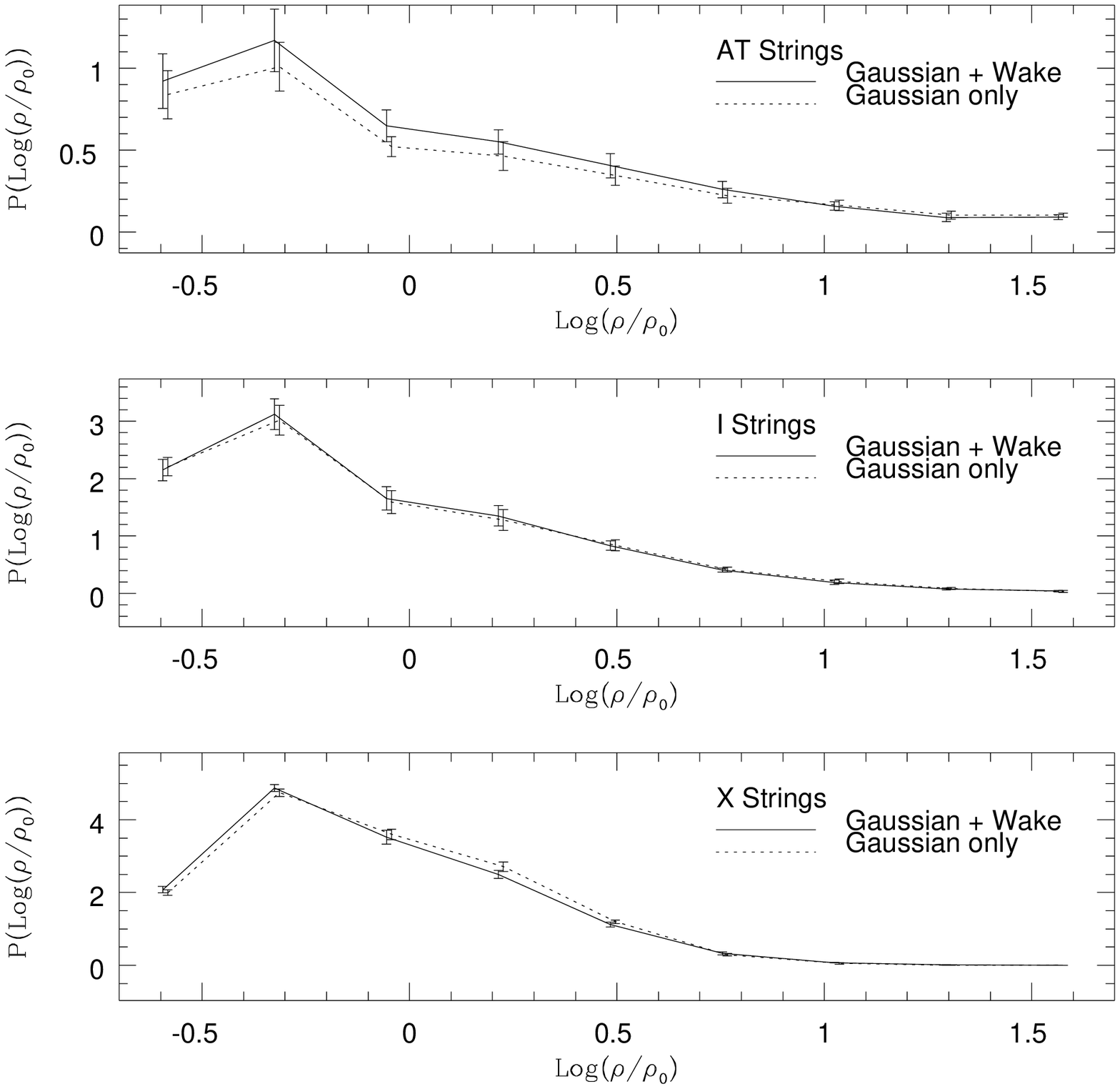,width=3.5in}}
\caption{Counts in cells distributions for X, I and AT boxes with and
without wakes. The boxes are $60$Mpc, $20$Mpc and $10$Mpc respectively. In
each case, the cell size is equal to the box size divided by $16$.}
\label{fig-cic16}
\end{figure}

Results for the counts in
cells statistic on our boxes for three different cell sizes
are shown in Figures \ref{fig-cic8}, \ref{fig-cic12} and
\ref{fig-cic16}. Here and throughout the paper, errorbars are shown at
the $2\sigma$ level. A
counts in cells analysis of the initial {\em linear} density contrast
would yield plenty of information about the string wake, since an
intrinsic effect of adding
in the wake is to add a large Non-Gaussian component to the counts in
cells distribution.
However, as the field undergoes gravitational collapse, this
information is lost, and the resulting probability distribution
has the same form as we would expect for an evolved Gaussian field. We
conclude that counts in cells is not a promising statistic with which
to identify string wakes.

\begin{figure}
\centerline{\psfig{file=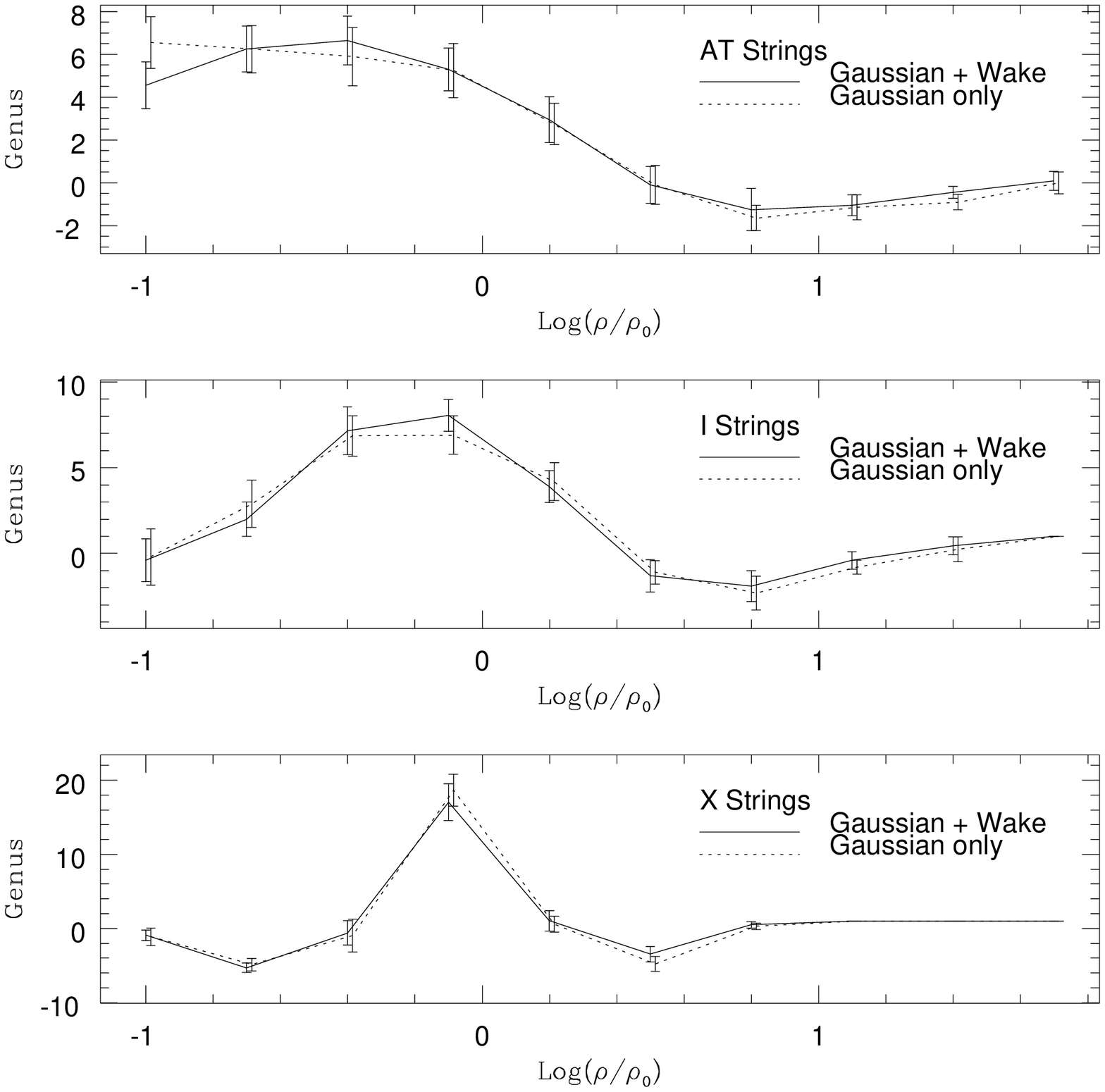,width=3.5in}}
\caption{Discrete genus curves for X, I and AT boxes with and
without wakes. The boxes are $60$Mpc, $20$Mpc and $10$Mpc respectively. In
each case, the cell size is equal to the box size divided by $8$.}
\label{fig-genus8}
\end{figure}

\begin{figure}
\centerline{\psfig{file=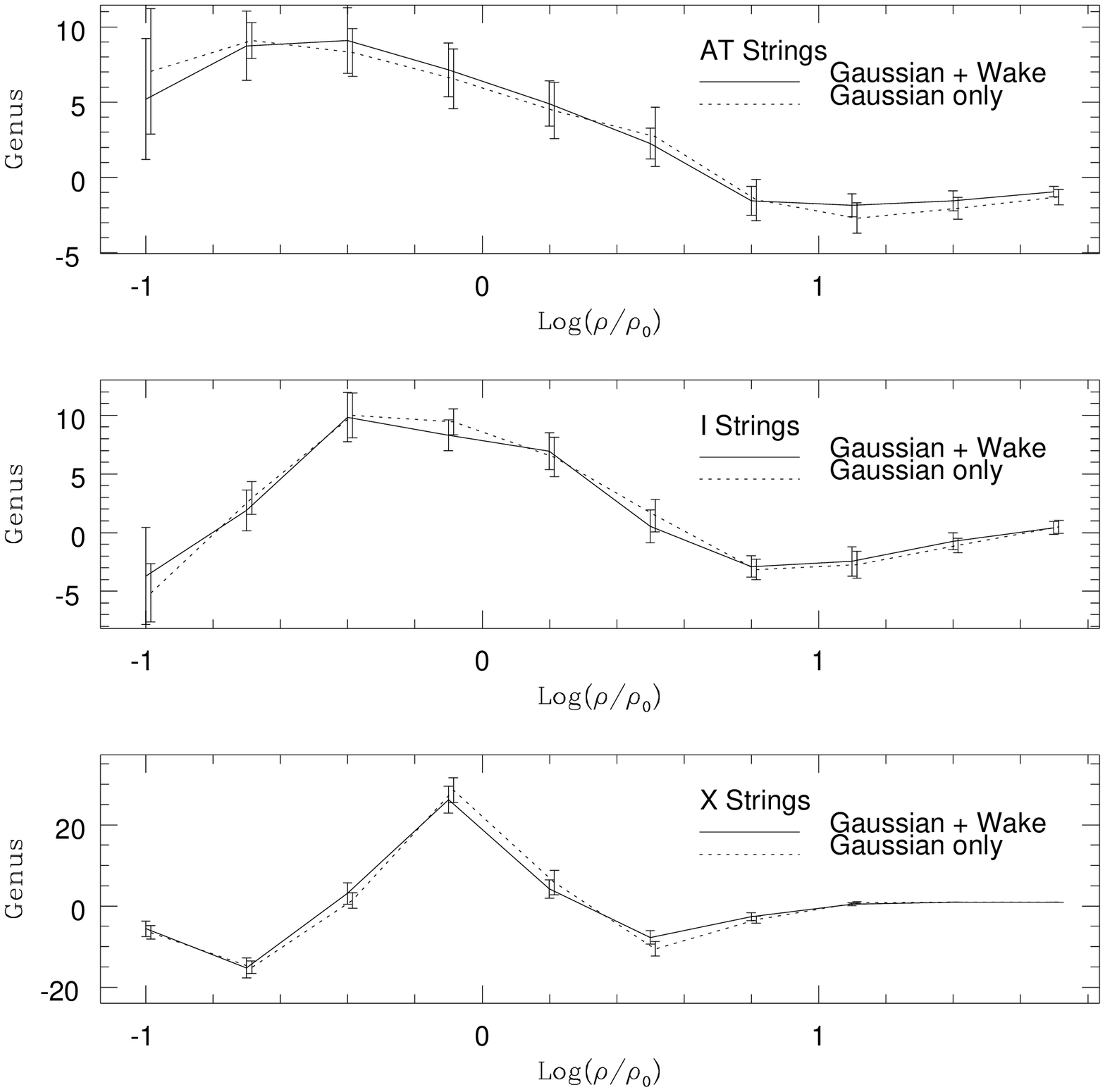,width=3.5in}}
\caption{Discrete genus curves for X, I and AT boxes with and
without wakes. The boxes are $60$Mpc, $20$Mpc and $10$Mpc respectively. In
each case, the cell size is equal to the box size divided by $12$.}
\label{fig-genus12}
\end{figure}

\begin{figure}
\centerline{\psfig{file=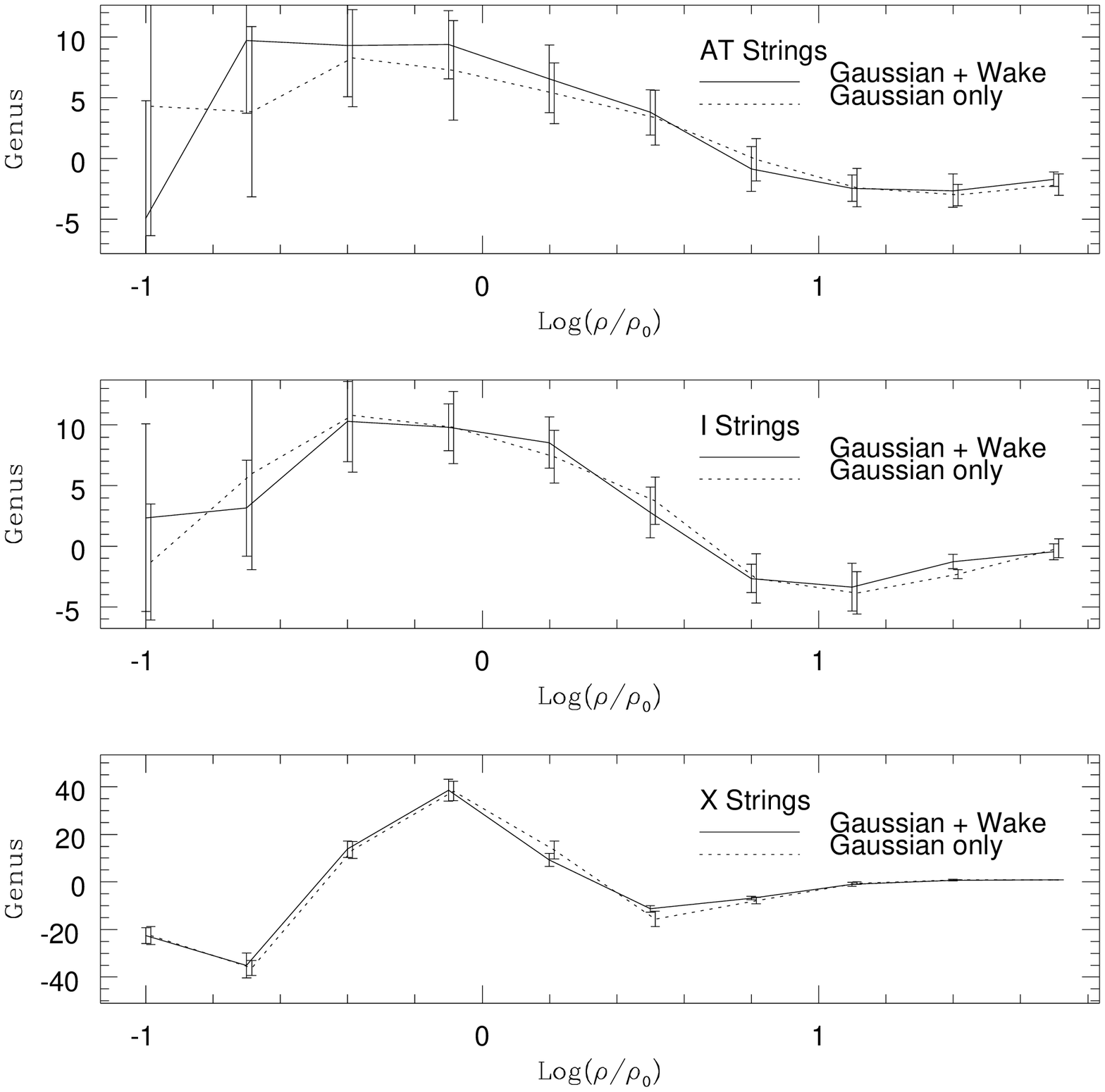,width=3.5in}}
\caption{Discrete genus curves for X, I and AT boxes with and
without wakes. The boxes are $60$Mpc, $20$Mpc and $10$Mpc respectively. In
each case, the cell size is equal to the box size divided by $16$.}
\label{fig-genus16}
\end{figure}

\begin{figure}
\centerline{\psfig{file=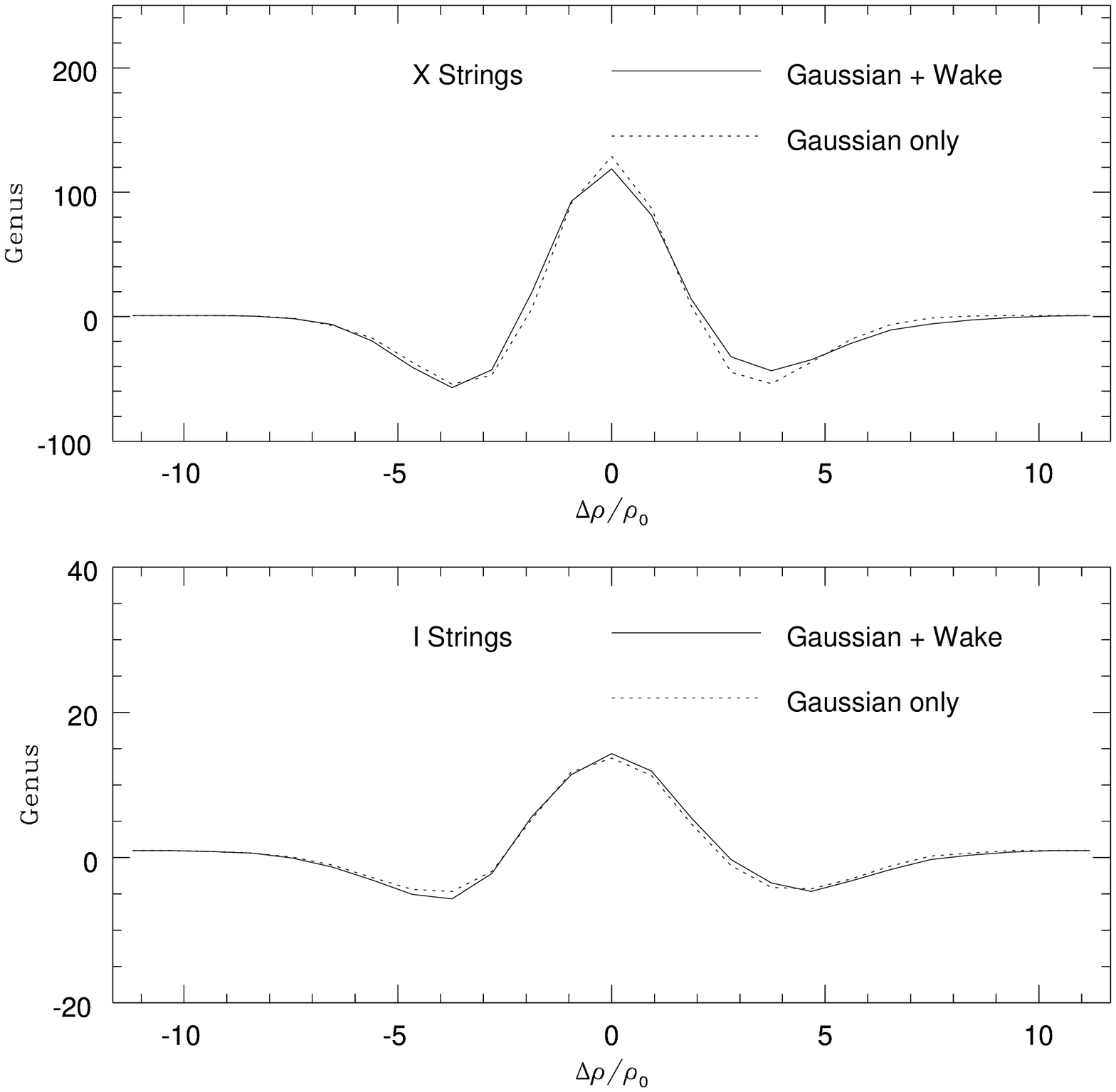,width=3.5in}}
\caption{Discrete genus curves for initial X and I boxes with and
without wakes. The boxes are $60$Mpc and $20$Mpc respectively. In
each case, the cell size is equal to the box size divided by $16$.
Isosurface cutoffs are normalized to the present day using linear gravity.}
\label{fig-initialgenus}
\end{figure}

Discrete genus curves \cite{brandenberger} for our
boxes are shown in
Figures \ref{fig-genus8}, \ref{fig-genus12} and \ref{fig-genus16}.
None of the string models produce a strong
signal in the curve.
The point is made even more strongly in
Figure~\ref{fig-initialgenus}, which shows genus curves for initial
density fields generated by I and X string. Even for the initial
density field, wakes only produce a tiny correction to the underlying
Gaussian form of the genus curve.
This suggests that the genus curve has little chance
of picking out string wakes today.

Brandenberger et al. \cite{brandenberger} argue on the basis of a different toy
model that wakes produce a genus curve and a counts in cells
distribution whose forms are qualitatively
different from those which arise in other theories of structure
formation. We feel that their wakes model tends to
exaggerate the sheet-like features in the density field.  On the other
hand, our toy model might be underestimating the presence of sheets,
at least in the X case (see Appendix \ref{sec-model}).  In the end,
a statistic will only be useful if it is effective in comparing
observations and more realistic realizations of theoretical models.
We expect that the enhanced sensitivity to wake-like features  of the
flatness statistic considered 
in the next subsection will make it a particularly useful tool.

\subsection{Looking for wakes via flatness}
\label{sec-flatstat}
In this subsection, we discuss statistics which might identify wakes
via their morphology. First, we consider an implementation of the
structure functions of Babul \& Starkman \cite{s123}. The structure
functions $S_1$, $S_2$ \& $S_3$ are a set of measures quantifying
the shape of an object. They take the value $1.0$ for a perfect
filament, plane or sphere respectively, and they fall off rapidly to
$0.0$ as the object deviates from each of these forms.

The starting point for these measures is the inertia tensor. For an
object consisting of a collection of point masses this is defined
as
\begin{equation}
    I_{ij}=M_{ij}-M_i M_j
\end{equation}
where
\begin{eqnarray}
        M_i&=&\frac{1}{M}\sum_{k} m^{(k)} r_i^{(k)}\\
        M_{ij}&=&\frac{1}{M}\sum_{k} m^{(k)} r_i^{(k)} r_j^{(k)}\\
        M&=&\sum_{k} m^{(k)}
\end{eqnarray}
and $m^{(k)}$ is the weight of the $k^{\rm th}$ point. Taking
$I_1$, $I_2$, $I_3$ to be the eigenvalues of $I_{ij}$ in decreasing
order of magnitude, $S_1$, $S_2$ \& $S_3$ are defined as
follows:
\begin{eqnarray}
    S_1&=& \sin \left[ \frac{\pi}{2} (1-\nu)^p \right]\\ 
    S_2&=& \sin \left[ \frac{\pi}{2} a(\mu,\nu)\right]\\
    S_3&=& \sin \left( \frac{\pi \mu} {2} \right)
\end{eqnarray}
where
\begin{eqnarray}
    \nu &=&\left(\frac{I_2}{I_1}\right)^{1/2}\\
    \mu &=&\left(\frac{I_3}{I_1}\right)^{1/2}\\
    p&=&\frac{\log 3} {\log 1.5}
\end{eqnarray}
and $a(\mu,\nu)$ is defined implicitly by
\begin{equation}
    \frac{\nu^2}{a^2} - \frac{\mu^2} {a^2 ( 1- \alpha a^{1/3} +
   \beta a^{2/3})} = 1
\end{equation}
with
\begin{eqnarray}
	\alpha&=&1.9\\
	\beta&=&-\frac{7}{8}9^{1/3}+\alpha 3^{1/3}
\end{eqnarray}
In addition, we consider a new flatness measure $F$, defined as
\begin{equation}
    F=\frac{ \sqrt{3} (I_2-I_1)(I_1^2+I_2^2+I_3^3)^{1/2} }
    {( I_1 I_1 + I_1 I_2 + I_1 I_3 +
    I_2 I_2 + I_2 I_3 + I_3 I_3)}  
\end{equation}
This measure is chosen to take the value 1.0 for a perfect plane,
but to fall away less rapidly to $0.0$ than $S_2$ as the object
deviates from planarity. In Appendix \ref{sec-shell} we compare the
values of $F$ and $S_2$ for sections of a spherical shell, which
may model realistic wakes better than a perfect plane.

In order to apply these statistics to a density field, we first need to
first identify objects in that field to measure the shapes of,
and then to employ some method
of averaging over the quantities obtained for these objects. Babul \& Starkman
\cite{s123} illustrate two important criteria for the identification
of objects.
\begin{itemize}
\item {\em Scale:} Firstly, they illustrate the strong dependence of 
$S_1$, $S_2$ and $S_3$ on the length scale on which objects are picked
out of the distribution.
\item {\em Thresholding:}
 Secondly, they show that the quantitative signal of an object can be
made stronger by ``thresholding'', that is, by throwing away those
parts of the matter distribution where the local density is below some cutoff
value\footnote{An alternative method for amplifying the signal of
an object is the minimal spanning tree. This technique has been
investigated by Pearson and Coles\cite{Pearson}. It would be
interesting to apply this technique to the density
fields considered in this paper.}.
\end{itemize}

 We adopt
the following scheme for identifying objects in our matter fields,
which  incorporates both of these criteria.

We divide our distribution into $i=1\ldots N$ cubes, $C_i$, of
side $R_{\rm smooth}$. We introduce thresholding by associating
a mass $m_i$ with the centre
of each cube, which is a function of the density inside
that cube $\rho_i$, and some cutoff density $\rho$, as follows.
\begin{equation}
    m_i= \rho \theta(\rho_i-\rho)
\end{equation}
where
\begin{equation}
    \theta(x)=\left\{
    \begin{array}{ll} 0 & \ldots x < 0 \\
                      1 & \ldots x\ge 0
    \end{array}
    \right.
\end{equation}
For each of the $M$ cubes $C_i$ for which the density $\rho_i$ is
above the threshold,
we construct an object $O_i$ which is the
set of all cubes $C_j$ contained within a cube of side $R$ centered on
$C_i$. (In order that we always deal with whole cubes, $R$ will be
some integer multiple of $R_{smooth}$).
For each object $O_i$ 
we work out the structure functions and the flatness statistic $S_1^i$,
$S_2^i$, $S_3^i$ and $F^i$. We then compute the average values
$S_1=\frac{1}{M}\sum_i S_1^i$ etc, providing a set of four 
numbers  $S_1$, $S_2$, $S_3$ and $F$
which quantify the morphology of the density field, as a function of
the three parameters $\rho$, $R$ and $R_{\rm smooth}$.

\begin{figure}
\centerline{\psfig{file=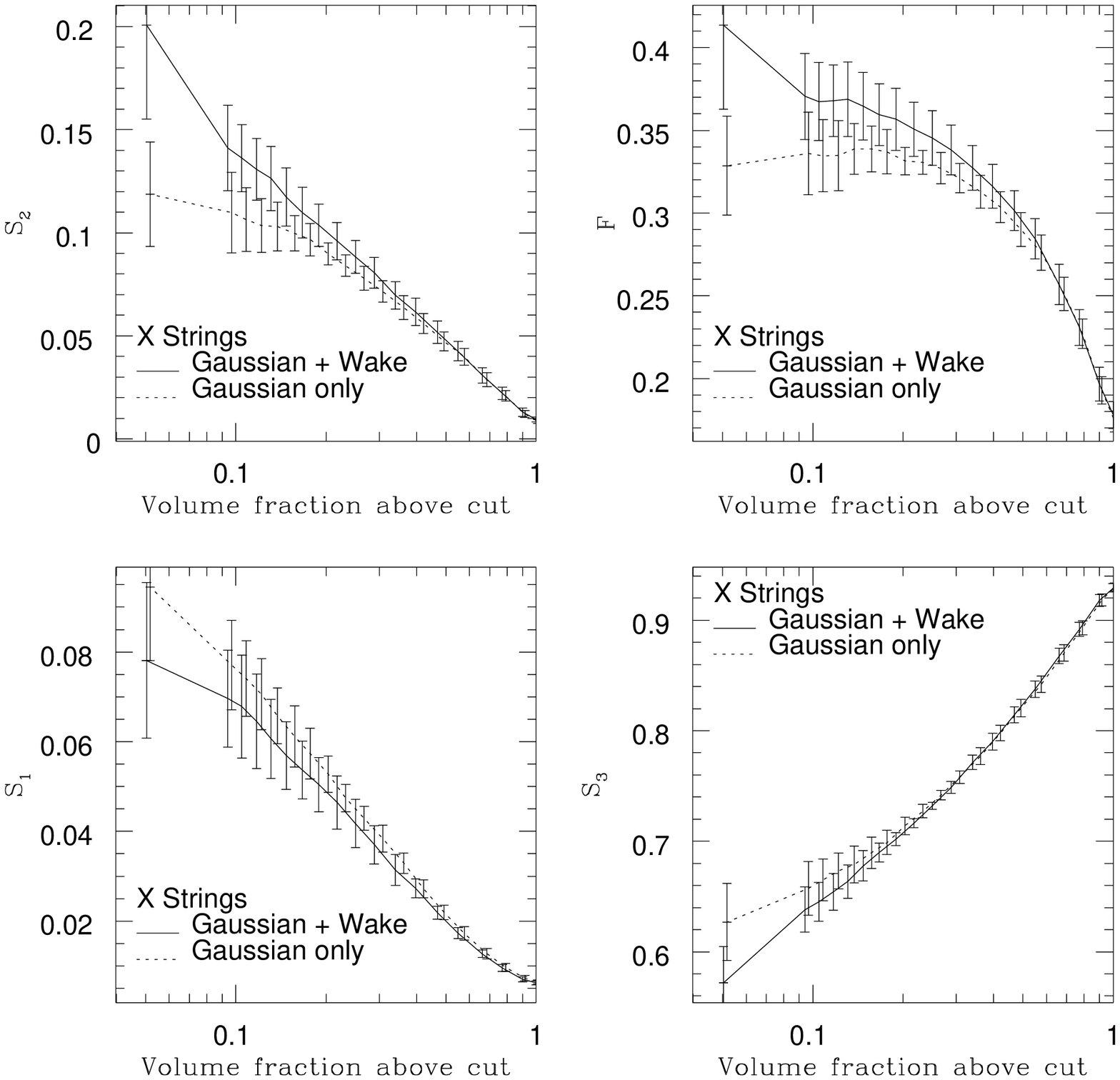,width=5.5in}}
\caption{Shapes curves for $60$Mpc boxes in the X string
model. $R$ is $\frac{3}{16}$ of the boxsize, and $R_{\rm smooth}$ is
$\frac{1}{16}$ of the boxsize.}
\label{fig-shapes3-3}
\end{figure}

\begin{figure}
\centerline{\psfig{file=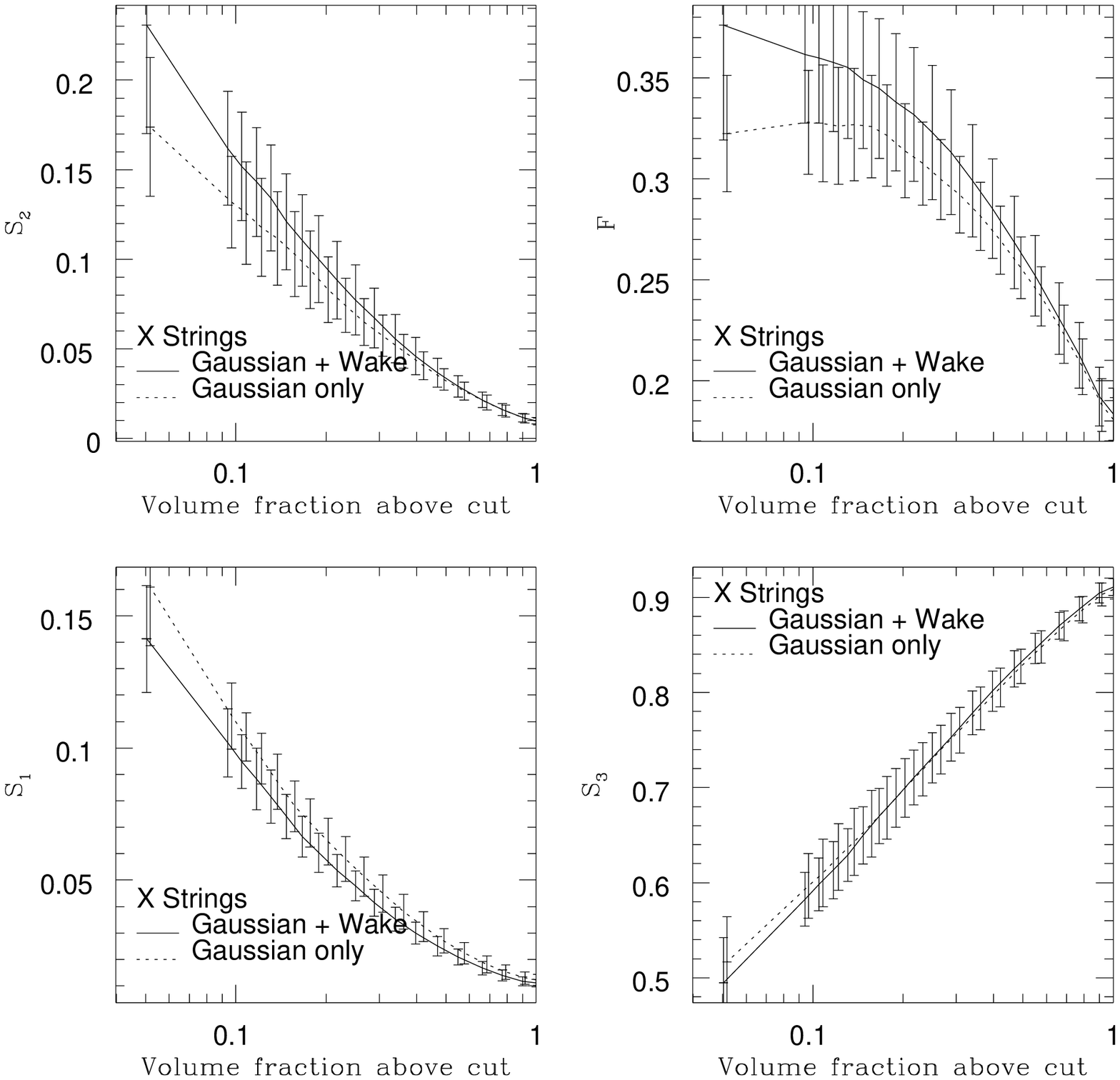,width=5.5in}}
\caption{Shapes curves for $60$Mpc boxes in the X string
model. $R$ is $\frac{7}{16}$ of the boxsize, and $R_{\rm smooth}$ is
$\frac{1}{16}$ of the boxsize.}
\label{fig-shapes3-7}
\end{figure}

\begin{figure}
\centerline{\psfig{file=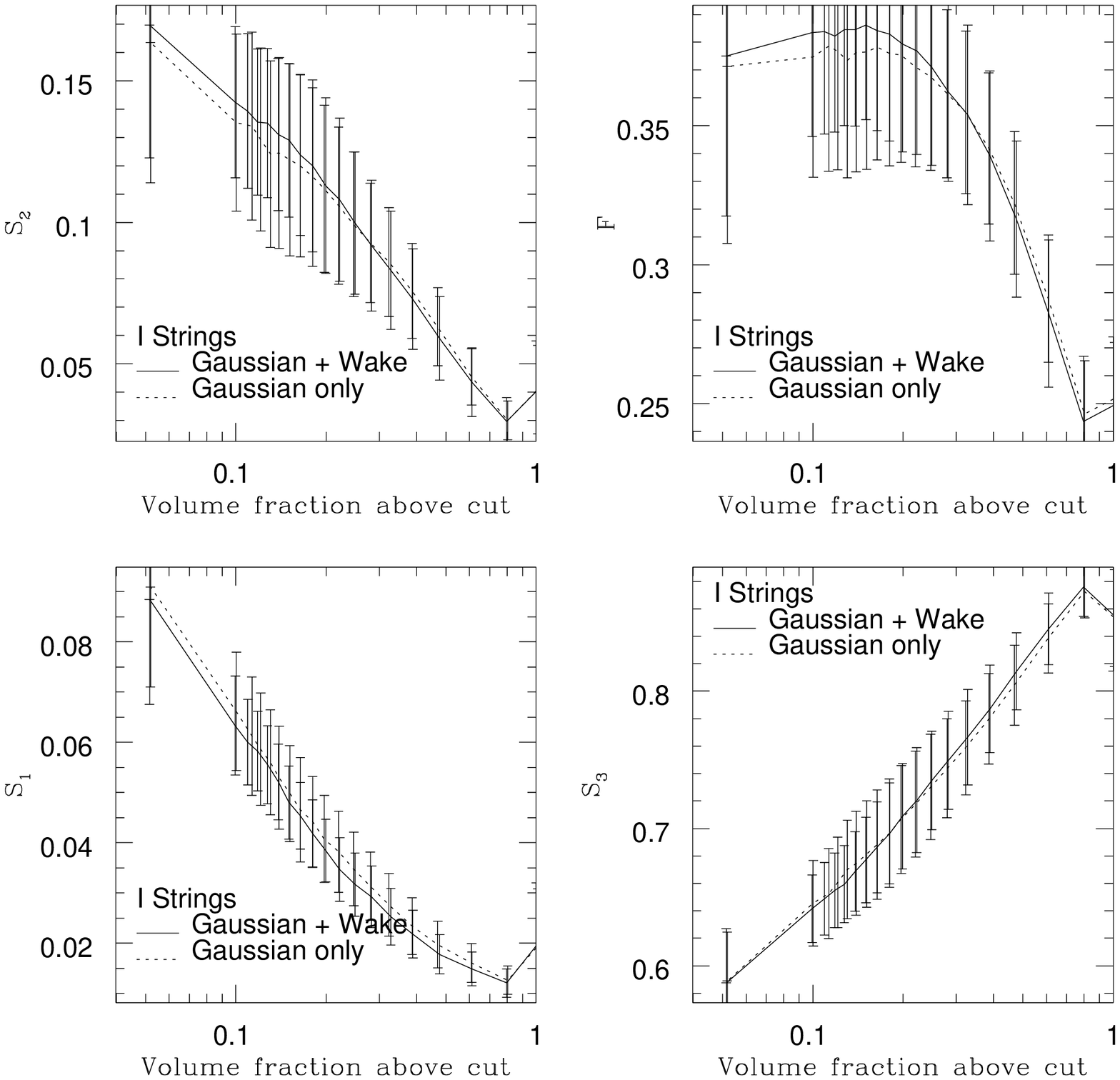,width=5.5in}}
\caption{Shapes curves for $20$Mpc boxes in the I string
model. $R$ is $\frac{3}{16}$ of the boxsize, and $R_{\rm smooth}$ is
$\frac{1}{16}$ of the boxsize.}
\label{fig-shapes2-3}
\end{figure}

In order to consider a wide range of parameter
space, we have developed a parallel computer code, and we present our
results as ``shapes curves'', showing the variation of $S_1$, $S_2$,
$S_3$ and $F$ with $\rho$, for given values of $R$. In Figures
\ref{fig-shapes3-3},\ref{fig-shapes3-7} and \ref{fig-shapes2-3}
we present a selection of results for the X and I string models, with various
choices for the parameters. All errorbars are shown at the $2\sigma$ level. 
In order to present our results in a
manner which is independent of the degree of non-linear evolution in
each of the models, for each cutoff $\rho$ we plot the
volume fraction $V$ of cubes
$C_i$ in the distribution for which $\rho_i>\rho$. We see that for a
suitable choice of parameters ($V \sim 0.05$, $R \sim 11$Mpc) we can
distinguish boxes with and without X string wakes at the $2\sigma$
level. However, none of the measures can distinguish boxes with and
without wakes in the I and AT models. We note that both $S_2$ and
$F$ are capable of picking out idealized flat X wakes at the
$2\sigma$ level, with $S_2$ performing slightly better. However, we
anticipate that the quantity F may perform better than $S_2$ when
picking out features which are not exactly flat (See Appendix
\ref{sec-shell}). As expected, the measures $S_1$ and $S_3$ which
quantify filamentarity and sphericity respectively are not useful for
distinguishing wakes.

We stress that the
model we have used to generate density fields is only a {\em toy}
model. We do not know whether string wakes will stand out more or less than
this in realistic models. If they stand out a little bit more, then
the statistic may be a useful tool for identifying them. If they stand
out less, then any statistic will have difficulty in distinguishing
them from the collapsed remnants of a Gaussian distribution. In general,
it seems that wakes which only just show up in initial density
fields (in particular, those that stand out on the linear gravity
criterion but fail to stand out on the velocity coherence
criterion -- see Appendix~\ref{sec-visibility}) tend to disappear
under non-linear gravitational evolution.
This observation may well be of importance when assessing the wakiness
of realistic models. 

Although we have tested the flatness statistic on a toy model for
cosmic string wakes, we also expect it to be useful for distinguishing
models with different Gaussian initial conditions.
As a density field undergoes gravitational collapse, it becomes harder
to retrieve information about the linear initial conditions which have
given rise to that collapse. This point is easily visualized in the
adhesion approximation, where degrees of freedom present in the initial
positions and velocities of particles are lost as the particles stick
together.
For a whole range of initial conditions, the first objects formed by
gravitational collapse are Zeldovich pancakes. It is possible that the
hard-to-retrieve information about linear initial conditions is best
encoded by these pancakes. The problems involved in
identifying and quantifying these pancakes are similar to
those involved in identifying X string wakes:
From the discussion in the previous
subsection, we expect that a flatness curve based on objects
contained by
isosurfaces will provide a good way of identifying these features.

\section{Discussion and Conclusions}
\label{sec-conclusion}
We have tried out various measures of the sheetiness of a
distribution of matter,
using a toy model of cosmic
string wakes in HDM.
In the most realistic version of our toy model (with an ``I'' string
network) we find that it is not possible to pick out single wakes.
However, our toy model is rather simplistic and it will be interesting
to see the extent to which improved models of the string network might
give a different result.  If anything, our model makes the wakes
unrealistically flat, and our expectation is that a more realistic
model will make the wakes less noticeable.

We have shown that wakes which fail to
satisfy our velocity coherence  criterion (see Appendix~\ref{sec-visibility})
at the linear stage
tend to stand out less as
a field undergoes non-linear gravitational evolution.   We  believe
that this observation will help simplify  assessing the wakiness of
more realistic string models.

We have also investigated an extreme (``X'' string) model, where the
wakes {\em
do} stand out strongly today. We have demonstrated that a good choice
of statistic is required to find the wakes, even in this case. Counts in
cells, the discrete genus curve, and some flatness statistics, do not
pick out the wakes well. However,
we have found that the ``flatness curve'', based on the structure
function $S_2$ of Babul \& Starkman,
produces a strong wakes signal.
The adhesion approximation we use for gravity  is known to exaggerate the
extent to which the Zeldovich pancakes are sheet-like. Thus our  work
might underestimate the differences between these pancakes and wakes.

We have shown that the flatness curve is 
a good statistic for
identifying X wakes (see Section~\ref{sec-flatstat}), and expect that it will also be useful
for identifying other types of sheet-like structure.
 We note that some properties of any initial density field
(including Gaussian fields) may be best encoded in the ``pancakey''
nature of the evolved distribution. Hence, the flatness
curve could well be a useful tool for
discriminating between many types of theories for the origin of large
scale structure.

\section{Acknowledgements}
We would like to thank Pedro Ferreira and Julian Borrill for helpful
discussions. We are also grateful for the use of the AP1000 at the
IC-Fujitsu Parallel Computing Research Centre.

\appendix
\section{The Model}
\label{sec-model}
Our picture of how cosmic string wakes are laid down can be visualized
in terms of the {\em one scale} model of string network evolution
\cite{AandT}\footnote{
Inadequacies in the one scale model have been exposed by high
resolution numerical simulations and a recently developed three scale
model attempts to address some of these inadequacies.  It would be
interesting to investigate how deviations from the one scale picture
could change our conclusions, but we expect the corrections to  be small.
}.
In the
one scale model, the statistics of the evolving network are specified
by one comoving scale $\xi$, a monotonically increasing
function of conformal time $\eta$, which characterizes the mean
inter-string separation, mean string curvature, and the mean step size of
the string walks. Suppose we lay down comoving boxes of side L at
random in the universe and want to work out how the string network
perturbs the matter in these boxes between its formation at time
$\eta_{\rm i}$, and the present day $\eta_0$. If we define $\eta_{\rm L}$ to
be the time
at which $\xi(\eta_{\rm L}) = L$, then we can separate the perturbation
history of the boxes into three epochs:

\begin{enumerate}
\item
\label{epoch1}
	$\eta_{\rm i} < \eta < \eta_{\rm L}$.
In this epoch, the string network describes
random walks with step size less than the box size. We approximate the
resulting perturbation as a Gaussian random
field, whose power spectrum may be
worked out using the method of Albrecht \& Stebbins
\cite{CDM}. (For the purposes of our toy model we have considered
perturbations in the HDM background
described by Albrecht \& Stebbins \cite{HDM}, with $h=1.0$,
$\sigma_8=1.0$ and no bias). Since we are only interested in
perturbations
seeded before time $\eta_{\rm L}$, we truncate the integration at
$\eta=\eta_{\rm L}$. So
\begin{equation}
\label{eqn-power}
   P(k)=16 \;\pi^2 \;\frac{(1+z_{\rm eq})^2}{(1+z_{\rm m})^2} \;\mu^2\;
        \int _{\eta_{\rm i}} ^{\eta_{\rm L}} | T(k,\eta) | ^2\;
F(k,\eta)\; {\rm d}\eta
\end{equation}
Here, $z_{\rm m}$ is the redshift at some time in the matter era to which we
can evolve the perturbations using linear gravity.

Clearly, as the time approaches $\eta_{\rm L}$, the perturbations produced
in this epoch become more and more sheet-like on  the scale of the
box. The question of whether
approximating all these perturbations as Gaussian random noise
is realistic can be assessed by
seeing how much the single wake we add stands out. For AT and I strings,
the wake stands out very little, so the approximation is
good.
For the X model, the wake stands out strongly,
so the approximation may be unreasonable.

\item
\label{epoch2}
        $\eta_{\rm L} < \eta < \eta_{\rm L} + \Delta \eta$.
Here $\xi$ is of order $L$, so strings
in the boxes look
roughly straight and move in roughly straight
lines.  Consequently, wakes laid down in any of the boxes
look roughly planar. The length of time $\Delta \eta$ required to
generate a box sized portion of one of these wakes is given by $\Delta
\eta = L/v_{\rm b}$,  where $v_{\rm b}$ is the macroscopic bulk velocity
of a
section of string in the network. Now, we can also estimate the {\em total}
area of string wakes generated in this time per unit volume of
universe. At any time, the length of string per unit volume of
universe is equal to $\xi^{-2}$. So in a time interval $d\eta$ the
area of wake traced out per unit volume is $ v_{\rm b}\; \xi^{-2} {\rm d}\eta$.
Consequently, the total wake area traced out per unit volume between
times $\eta_{\rm L}$ and $\eta_{\rm L} + \Delta \eta$ is given by
\begin{equation}
	a_2 = v_{\rm b} \int _{\eta_{\rm L}} ^{\eta_{\rm L} + \Delta \eta} \;
	\frac {{\rm d}\eta} {\xi^2}
\end{equation}
Consider the X string model, where $\xi(\eta)=\eta$. Then
\begin{equation}
   a_2 = v_{\rm b} \left( \frac {1} {\eta_{\rm L}} - \frac{1}
{\eta_{\rm L} + \Delta \eta}
	\right)
     = \frac {v_{\rm b}} {L (v_{\rm b}+1)}
\end{equation}
Every time a box hits a wake, the area of wake per unit volume
contained in the box is of order $L^2/L^3 = L^{-1}$. Let the fraction
of boxes hitting wakes laid down in this epoch be $p_2$. Then
equating string area per unit volume over an ensemble of boxes we have
\begin{equation}
   \frac {p_2} {L} = \frac{v_{\rm b}} {L(v_{\rm b}+1)}
\end{equation}
Hence we can work out
\begin{equation}
   p_2 = \frac {v_{\rm b}} {1+v_{\rm b}}
\end{equation}
(The results for I and AT models will be similar). Then the
perturbations in this epoch are
well approximated as follows
\begin{enumerate}
\item
   A fraction $p_2$ of the boxes are perturbed by a single string,
which by an appropriate choice of axes we may take to have moved in a
straight line through the middle.
To work out the density
perturbation we
simply calculate $\theta_+(k,\eta)$ for a straight string moving in a
straight line through the middle of the box, and
substitute into equation (2) of \cite{CDM}. We use $v_{\rm b} = 0.3$ for the
AT and X models, and $v_{\rm b}=0.15$ for the I model. The difference reflects
the enhanced small scale structure in the I model.
\item
   A fraction $1-p_2$ of the boxes do not undergo any perturbations in
this epoch.
\end{enumerate}

\item
	$\eta_{\rm L}+\Delta \eta < \eta < \eta_0$.
The area of wakes per unit volume generated in this epoch is given by
\begin{equation}
a_3 =v_{\rm b}\; \int _{\eta_{\rm L} + \Delta \eta} ^{\eta_0}\; \frac
{{\rm d} \eta} { \xi^2}
\end{equation}
which for $\eta_0 \gg \eta_{\rm L}$ in the X model works out to
\begin{equation}
a_3 =  \frac {v_{\rm b}^2} { L(v_{\rm b} +1)}
\end{equation}
$\xi$ is now greater than $L$, so wakes laid
down in any of the boxes look very planar. Hence, the area of wake per
unit volume in
any box which hits a wake is approximately $L^{-1}$. Let the fraction of
boxes hitting wakes laid down in this epoch be $p_3$. Doing the same
calculation as in the last epoch, we see that $p_3 = \frac {v_{\rm b}^2} {
(v_{\rm b} +1)}$, which is always less than $0.5$, and may be as low as
$.019$ if $v_{\rm b} = 0.15$. So the majority of boxes, a fraction $1-p_3$,
undergo no perturbations in this epoch. We will consider only these
boxes. (Again, results will be similar for the AT and I models).

\end{enumerate}

This picture provides a simple way of working out what the wakes
laid down at time $\eta_{\rm L}$ will look like today:
Provided $\eta_{\rm L}$ lies in the regime where linear gravity is
a good approximation, we can choose some later time $\eta_{\rm M}$ in
the matter
dominated era of the universe, at which linear gravity is still a good
approximation. We can work out perturbations induced during
Epoch~\ref{epoch1} at
time $\eta_{\rm M}$ by computing a realization of
the power spectrum in Equation
(\ref{eqn-power}). We can work out perturbations induced during
Epoch~\ref{epoch2}
using Equation (2) of \cite{CDM}. Up to time $\eta_{\rm M}$ gravity is still
linear so the
combined effect of these perturbations can be obtained by simply adding
them together. Since no further perturbations occur in
the chosen boxes, we can evolve these initial conditions to the present day
using standard N-body techniques\footnote{For speed, we perform the
same evolution using the adhesion approximation. For our statistics we
evolved 10 realizations of each model, with $32^3$ particles on a
$32^3$ grid. The adhesion approximation has one free parameter, the
viscosity, which we set to be $(2$ grid units$)^2$.}.

As a final point, we address a technical question connected with applying
the adhesion approximation to a model with HDM.
In our HDM scenario, the evolution of density on any scale up to a
redshift of 400 is described very
well by linear gravity.  This is because up to this time the density
contrast
averaged over an arbitrarily small
volume is always less than or of order 0.01. After this time, the
effects of non-linear gravity
start to be important. Free streaming by neutrinos does not
affect this non-linear
evolution on any interesting scales, because the typical distance
moved by a neutrino between this time and today is of order 0.2Mpc.
So we only need to take HDM into account in the linear regime, and we can use
standard cold N-body techniques to carry out the subsequent non-linear
evolution.

\section{A wake visibility criterion based on velocity coherence}
\label{sec-visibility}
In Section~\ref{sec-qualitative} we described a criterion based on linear
gravity for testing whether cosmic string wakes stand out. Now, a
simple alternative to linear gravity is given by the Zeldovich
approximation,
which gives a more realistic
picture of the evolution of a distribution into the non linear regime.
In the Zeldovich
 approximation, the evolution of the matter is
determined by the {\em velocities} of test
particles laid down in the initial density field. We might hope that
an improvement on the linear gravity
wakiness criterion would be provided by a similar ``velocity coherence''
criterion. Here, we ask whether the dominant contribution to the mean
velocity
of a sphere of radius $R$ sitting on the edge of a wake comes from the
wake ($v_{\rm W}$) or from the Gaussian background ($v_{\rm rms}$). If
$v_{\rm W}
\succeq v_{\rm rms}$ then the wake may be said to stand out.
We have
\begin{eqnarray}
   v_{\rm W} &=& 8 \;\mu\; \Sigma\; (1+z_{\rm eq})\; \times
\nonumber\\
&&
              \int _0 ^\infty \; {\rm d}k \frac {\sin kR}{k} \;
	      w(kR)\; T(k,\eta_{\rm W})\\
   v_{\rm rms}^2& =& 4 \;\pi\;
              \int _0 ^\infty \; {\rm d}k \;
	      |w(kR)|^2\; P(k)
\end{eqnarray}
where $w(x)$ is the spherical window function used by Albrecht \&
Stebbins \cite{HDM} and $\eta_{\rm W}$ is the time at
which the wake in question has been laid down.

\begin{figure}
\centerline{\psfig{file=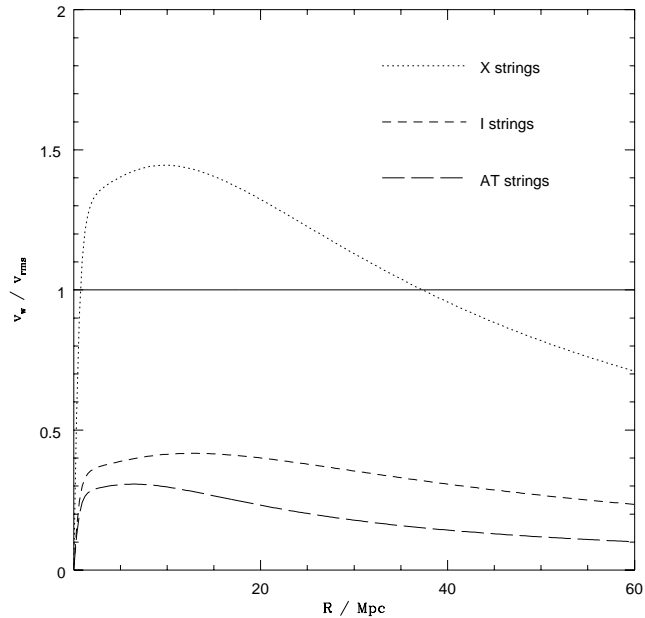,width=3.5in}}
\caption{Graph showing the ratio $v_{\rm W}/v_{\rm rms}$ as a function
of sphere radius for spheres
laid down on the edge of ``maximal'' wakes in the AT, I and X string models.}
\label{fig-velocities3}
\end{figure}

\begin{figure}
\centerline{\psfig{file=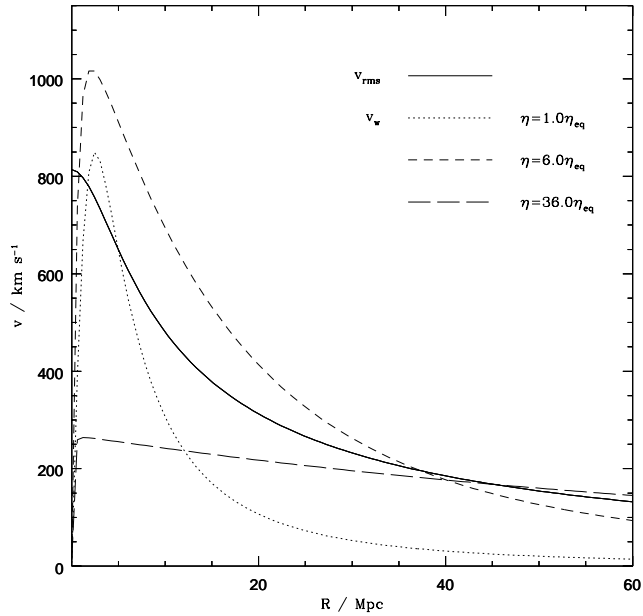,width=3.5in}}
\caption{Graph showing $v_{\rm rms}$ and $v_{\rm W}$ as a function
of sphere radius for spheres
laid down on the edge of wakes in the X string model. Each dotted
curve is for a wake laid down at a particular time. The velocities are
normalized to the present day using linear gravity.}
\label{fig-velocitiesX}
\end{figure}

Results for this criteria for maximal wakes in the three models are
given in Figure \ref{fig-velocities3}.
Results for wakes laid down
over a range of times in the X model are shown in Figure
\ref{fig-velocitiesX}.
 It is seen that only the X wakes satisfy the
velocity coherence criterion. That is, only X wakes dominate the bulk
motions of nearby matter. Because of this, they will tend to stand out
more and more as they undergo non-linear gravitational evolution. In a
string model producing wakes of
this type, the universe should
contain high density sheets which directly track the motions of
strings. On
the other hand, wakes which do not satisfy the velocity coherence
criterion will tend to
break up and become less visible through non-linear evolution.
These facts are borne out by an examination of the pictures in
Section~\ref{sec-eyeball}.

We expect that the velocity coherence criterion will provide a good
test of whether wakes present in linear density fields will be
visible in the universe today.

\section{Flatness of sections of a spherical shell}
\label{sec-shell}
\begin{figure}
\centerline{\psfig{file=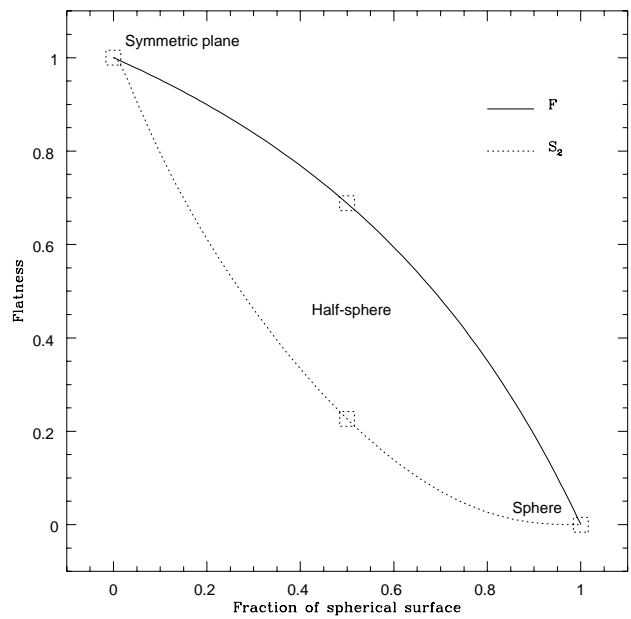,width=3.5in}}
\caption{Graph showing the flatness of various fractions of a
spherical shell.}
\label{fig-flatness_shells}
\end{figure}

Strings are curved, and they move in curved paths. In the one scale
model, the degree of curvature is characterized by by a radius of
curvature of order $\xi$.
It is often useful to imagine that the sheets they trace out are
spherical shells of radius $\xi$ \cite{stebbins-cam}. In order for our
flatness statistic
to pick out wakes, it must be sensitive to the flatness of sections of
these shells.
Figure~\ref{fig-flatness_shells} shows the value of the flatness
quantities $F$ and $S_2$
described in \ref{sec-flatstat} for various sections of a
spherical shell. Each section is constructed by cutting the sphere
with a plane, and only considering that part of the sphere lying on
one side of the cut. We quantify the shell fraction as the ratio of
the area of the section to the area of the sphere.
We observe that $F \succeq 0.7$ for sections ranging from the planar
limit to a half-sphere, whereas $S_2$ falls to $0.2$ for a
half-sphere. We anticipate that $F$ may be more sensitive than $S_2$
for identifying wakes and other sheet-like features which do not
start off exactly flat.


\begin{thebibliography}{}

\bibitem{CDM}
Albrecht A., Stebbins A., 1992a, Phys. Rev. Lett., 68,
2121

\bibitem{HDM} Albrecht
A., Stebbins A., 1992b, Phys. Rev. Lett., 69, 2615

\bibitem{AandT} Albrecht
A., Turok N., 1989, Phys. Rev. D, 40, 973

\bibitem{s123} Babul A.
Starkman G.D., 1992, ApJ, 401, 28

\bibitem
{brandenberger} Brandenberger R.H., Kaplan D.M., Ramsey S.A.,
1993, astro-ph/9310004

\bibitem{gott1}
Gott J., Melott A., Dickinson M., 1986, ApJ, 306, 341

\bibitem{gott2}
Gott J., Weinberg D., Melott A., 1987, ApJ, 319, 1

\bibitem{Pearson} Pearson
R.C., Coles P., 1995, MNRAS, 272, 231

\bibitem{times} Perivolaropoulos L., Brandenberger R.H.,
Stebbins A., 1990, Phys. Rev. D, 41, 1764

\bibitem{salsaw}
Salsaw W., 1989, ApJ, 341, 588

\bibitem{stebbins-cam}
 Stebbins A., 1990, in Gibbons G., Hawking S., Vachaspati T., eds, The
Formation and Evolution
of Cosmic Strings,
Cambridge University Press, Cambridge, p. 503

\bibitem{visniac} Vishniac E., 1986, in
Kolb E.W., Turner M.S., Lindley D., Olive K., Seckel D., eds, Inner
Space Outer Space. University of Chicago Press, p. 190

\bibitem{adhesion} Weinberg
D.H., Gunn J.E., 1990, MNRAS, 247, 260


\end{thebibliography}
\end{document}